\documentclass[aps,pra,twocolumn,showpacs, superscriptaddress]{revtex4-1}

\usepackage{graphicx}
\usepackage{amsmath}
\usepackage{graphicx}
\usepackage{amssymb}
\usepackage{epstopdf}
\usepackage{appendix}
\usepackage{enumerate}
\usepackage{hyperref}

\usepackage{color}

\newcommand{\nc}{\newcommand}
\nc{\eq}{\begin{equation}}
\nc{\eeq}{\end{equation}}
\nc{\eqa}{\begin{eqnarray}}
\nc{\eeqa}{\end{eqnarray}}
\nc{\ar}{\begin{array}}
\nc{\ear}{\end{array}}
\nc{\bfig}{\begin{figure}}
\nc{\efig}{\end{figure}}
\nc{\dg}{\dagger}
\nc{\sx}{\sigma_x}
\nc{\sy}{\sigma_y}
\nc{\sz}{\sigma_z}
\nc{\spl}{\sigma_+}
\nc{\sm}{\sigma_-}
\nc{\nn}{\nonumber}
\nc{\noi}{\noindent}
\nc{\adg}{a^{\dg}}
\nc{\kvec}{\mathbf{k}}

\def\bra#1{\mathinner{\langle{#1}|}}
\def\ket#1{\mathinner{|{#1}\rangle}}

\begin{document}

%\preprint{}

\title{Entanglement control via reservoir engineering in ultracold atomic gases}



\author{S. McEndoo}
\email[]{sm636@hw.ac.uc}
\homepage[]{www.openquantum.co.uk}
\affiliation{Turku Center for Quantum Physics, Department of Physics and Astronomy, University of Turku, FIN-20014 Turku, Finland}
\affiliation{SUPA, EPS/Physics, Heriot-Watt University, Edinburgh, EH144AS, UK}
\author{P. Haikka}
\affiliation{Turku Center for Quantum Physics, Department of Physics and Astronomy, University of Turku, FIN-20014 Turku, Finland}
\author{G. De Chiara}
\affiliation{Centre for Theoretical Atomic, Molecular and Optical Physics, Queen's University Belfast, Belfast BT7 1NN, United Kingdom}
\author{G. M. Palma}
\affiliation{NEST Istituto Nanoscienze-CNR and Dipartimento di Fisica, Universit\`a degli Studi di Palermo, via Archirafi 36, I-90123 Palermo, Italy}
\author{S. Maniscalco} 
\affiliation{Turku Center for Quantum Physics, Department of Physics and Astronomy, University of Turku, FIN-20014 Turku, Finland}
\affiliation{SUPA, EPS/Physics, Heriot-Watt University, Edinburgh, EH144AS, UK}
\email[]{smanis@utu.fi} \homepage[]{www.openquantum.co.uk}


\date{\today}

\begin{abstract}
We study the entanglement of two impurity qubits immersed in a Bose-Einstein condensate (BEC) reservoir. This open quantum system is particularly interesting because the reservoir and system parameters are easily controllable and the reduced dynamics is highly non-Markovian. We show how the model allows for interpolation between a common dephasing scenario and an independent dephasing scenario by simply modifying the wavelength of the superlattice superposed to the BEC, and how this influences the dynamical properties of the impurities.
We demonstrate the existence of very rich entanglement dynamics correspondent to different values of reservoir parameters, including phenomena such as entanglement trapping, entanglement sudden death, revivals of entanglement, and BEC-mediated entanglement generation. In the spirit of reservoir engineering, we present the optimal BEC parameters for entanglement generation and trapping, showing the key role of the ultracold gas interactions. 

\end{abstract}

%\keywords{}

\maketitle

\section{Introduction}
Ultracold gases have recently emerged as an exciting playground for the simulation of complex many-body systems. The great level of control and cooling over neutral atoms in an optical lattice has opened new avenues for the study of quantum magnetism, disordered systems, long range interactions, supersolidity, and the effect of non-Abelian fields, just to mention a few examples (see, for example, Refs.~\cite{bloch_review,anna_book}). A particularly rich area of research at the point between ultracold atoms, open quantum systems and quantum information is the study of quantum reservoir engineering.

Mesoscopic quantum systems can be viewed as a special form of environment: due to their low temperature and relatively small size, their quantum coherence leads to important effects. A Bose-Einstein condensate (BEC) is a typical example. Previous works have  focused on the interaction of localised impurities immersed in a BEC \cite{AQD1}; others have studied the case of a lattice of impurities interacting with a BEC which leads to emission of phonons and dissipation for the impurities \cite{AQD2}. The collective dephasing of a two spatial states impurity was considered in \cite{gabriele}, and the dynamics of a single impurity in a BEC was recently investigated in \cite{us,Mulansky,peotta}. Moreover, experiments realising the dynamics of a single impurity immersed in a BEC \cite{Will,widera} have recently been reported.
All these studies and experimental progress open the way to applications in quantum information processing that were, until recently, unrealistic. 
%
%Here we study for the first time the entanglement dynamics of two impurity atoms immersed in a bosonic gas. Not only we show in which regime the impurity-impurity entanglement can be preserved from decoherence induced by the BEC but also the conditions under which the bosonic environment is able to create entanglement between the two, initially separable, atomic impurities. 
%{\color{red} 

In general, the presence of the ultracold gas surrounding the impurities gives rise to non-dissipative decoherence due to collisions between the BEC atoms and the impurity atoms. The literature on both entanglement decay and entanglement generation in the presence of dephasing environments has focused on  certain paradigmatic models of environment with specific types of spectral densities  \cite{dephasing, weiss,braun}. In this paper we investigate for the first time a situation for which the entanglement dynamics can be modified by changing both the form of the spectrum and the distance between the impurities. In fact, in our system a modification of the BEC scattering length is reflected in the form of the reservoir frequency spectrum, leading to a change in its Ohmicity character \cite{us}. As a consequence we can investigate the optimal reservoir parameters to preserve both entanglement and discord as well as the optimal conditions for entanglement generation from a separable state. 

From a fundamental point of view our results pave the way to new experiments on basic aspects of open quantum systems. Indeed, they provide a physical example of an experimentally realisable bipartite open quantum system whose exact, and therefore non-Markovian, dynamics is described by a time-local master equation. This is important because, together with optical implementations \cite{expnonM}, these systems could be used as testbeds for experimental verification of fundamental theorems of open quantum systems, thus giving an essential contribution to the thriving field of non-Markovian quantum dynamics \cite{nonmark}. More specifically, the two-impurities in BEC we discuss in this paper could be used to verify the simplest nontrivial generalisation of the Lindblad-Gorini-Kossakowski-Sudarshan (LGKS) theorem for commutative dynamical generators, as we briefly discuss in Sec. II.

The structure of the paper is the following. In Sec. II we recall the microscopic derivation of the effective pure dephasing model, we derive the master equation for the two impurities, and discuss its interest for fundamental tests of open quantum systems dynamics. In Sec. III we discuss the dynamics of both entanglement and quantum discord for different reservoir parameters. In Sec. IV we show the onset of entanglement generation and its dependence on the distance between the qubits and on the scattering length. Finally, in Sec. V we present conclusions.   

%}

%%%%%%%%%%%%%%%%%%%%%
\begin{figure}[ht]
\includegraphics[scale=0.5,angle=90]{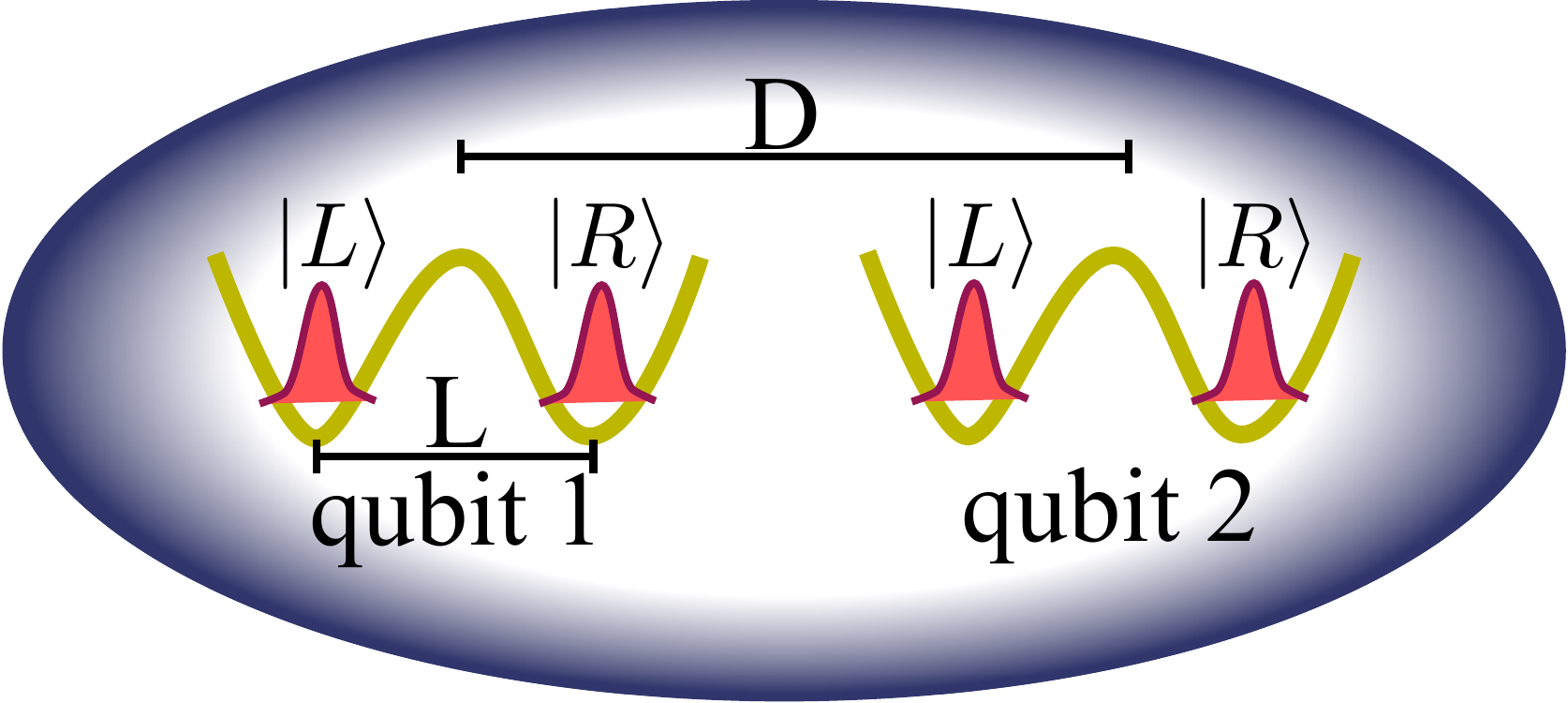}
\caption{(Color online) Schematic view of the proposed setup. Two impurities are each separately trapped in a deep double well potential, such that tunnelling between wells is suppressed. In this regime each atom occupies the two states localized in each well. The two impurities are immersed in a three-dimensional BEC with which they interact via cold collisions. We denote by $L$ and $D$ the separation between the two wells in an impurity and the separation between the two double-wells, respectively.}
\label{fig:setup}
\end{figure}

\section{The model} 
The system we consider consists of two impurity atoms immersed in a Bose-Einstein condensate reservoir trapped by a large harmonic potential. In contrast to similar proposals in which atoms were confined in a tight harmonic trap \cite{AQD1,AQD2}, in the present model, sketched in Fig.~\ref{fig:setup}, each atom is trapped in a double well potential. The two minima are separated by a distance $L$, with the double wells separated by distance $D$. Our qubit consists of the presence of the atom in the left or right well of the double well. The combined system is governed by the following Hamiltonian~\cite{gabriele}
\eq
\label{eq:totalH}
	\hat{H} = \hat{H}_A + \hat{H}_B + \hat{H}_{AB},
\eeq
where
\eq
	\hat{H}_A = \int d\mathbf{x}  \hat{\Psi}^\dagger({\bf x}) \left[ \frac{{\bf p}^2_A}{2 m_A} + V_A({\bf x}) \right] \hat{\Psi}({\bf x})
\eeq
is the Hamiltonian of the atom impurity with mass $m_A$, where $\hat{\Psi}({\bf x})$ is the corresponding atomic field operator~\cite{note}, $V_A({\bf x})$ is the double well potential formed by the optical lattice,
\eq
	\hat{H}_B = \int d\mathbf{x} \hat{\Phi}^\dagger({\bf x}) \left[  \frac{{\bf p}^2_B}{2 m_B} + V_B({\bf x}) + \frac{g_B}{2}  \hat{\Phi}^\dagger({\bf x})\hat{\Phi}({\bf x})  \right] \hat{\Phi}^({\bf x})
\eeq
is the Hamiltonian for the BEC atoms of mass $m_B$, $\hat{\Phi}({\bf x})$ is the corresponding field operator satisfying bosonic commutation relations, $V_B({\bf x})$ is the BEC trapping potential which we assume to be constant and $g_B = 4\pi \hbar^2a_B/m_B$ is the boson-boson coupling constant, with $a_B$ the BEC scattering length. We denote by 
\eq
	\hat{H}_{AB} = g_{AB} \int d\mathbf{x} \hat{\Psi}^\dagger({\bf x})   \hat{\Phi}^\dagger({\bf x}) \hat{\Phi}({\bf x})  \hat{\Psi}({\bf x})
\eeq
 the interaction Hamiltonian of the impurity atoms with the condensate, and $g_{AB} = 2\pi\hbar^2 a_{AB}/m_{AB}$ is the coupling between the impurity atoms and the condensate gas. We defined the reduced mass $m_{AB}=m_Am_B/(m_A+m_B)$ and the scattering length $a_{AB}$ corresponding to impurity-boson collisions. We now make a number of approximations leading to a simplified model. First we assume deep double wells, meaning that the impurity atoms can only be found in the lowest Wannier states, $\ket{L}$ and $\ket{R}$, of the double wells corresponding to the atom localised in the left and right well respectively and we neglect tunnelling across the barrier separating the two wells. This allows one to rewrite Hamiltonian \eqref{eq:totalH} using the pseudo-spin operator $\hat\sigma_z=\ket{L}\bra{L}-\ket{R}\bra{R}$. The second approximation valid for a weakly interacting Bose gas, consists of expanding the field operator  as
 \begin{equation}
\hat\Phi({\bf x})=\sqrt{N_0}\Phi_0({\bf x})+\sum_\kvec \left[u_\kvec({\bf x}) \hat c_\kvec - v_\kvec^*({\bf x}) \hat c_\kvec^\dagger\right]
\end{equation}
where $\Phi_0({\bf x})$ is the classical condensate wave function and the operators $\hat c_k$ correspond to the Bogoliubov modes of the BEC (see for example \cite{BEC}).
Using this assumption, and following the derivation of Ref.~\cite{gabriele} we arrive to an independent boson model with an effective Hamiltonian:
\begin{widetext}
\begin{eqnarray}\label{htot}
 \hat{H} = \hat{H}_{Bog}
+ \frac{\hbar}{2} \sum_\kvec \left\{ \left( \sum_{i=1,2} ( \Omega_{R,\kvec}^i -\Omega_{L,\kvec}^i) \hat{\sigma}_z^i +  \sum_{i=1,2}( \Omega_{R,\kvec}^i + \Omega_{L,\kvec}^i) \right) \hat{c}_\kvec
+ \textrm{h.c.}\right\}
\end{eqnarray}
\end{widetext}
where
\begin{equation}
\hat{H}_{Bog}=\sum_\kvec E_k \hat{c}^\dagger_\kvec \hat{c}_\kvec
\end{equation}
is the Bogoliubov Hamiltonian and $E_k$ is the corresponding energy. The constants $\Omega_{R,\kvec}^i$ and $\Omega_{L,\kvec}^{i}$ are the coupling frequencies of the left and right localised states of qubit $i$ respectively to the $\kvec$-th Bogoliubov mode, and $h.c$ is the Hermitian conjugate. The form of the constants $\Omega$ depends on the overlap integrals of the impurities Wannier functions with the wave functions associated to the Bogoliubov modes. Their complete form and their derivation can be found in \cite{gabriele}. We assume a constant BEC wave function $\Phi_0$ that is not modified by the presence of the impurities. This means that the terms of the Hamiltonian responsible for the Gross-Pitaevskii Equation do not enter into Eq.~\ref{htot}. Finally $\hat\sigma_z^i$ is the Pauli operator of qubit $i$.
Further details about possible experimental realisations are given in Appendix \ref{app}.

We assume an initially factorized state of the environment and the two qubits:
\begin{equation}
\varrho(0) = \rho(0)\otimes\tau_B
\end{equation}
where $\rho(0)$ is the initial (possibly correlated) density matrix of the two qubits that we will specify later and $\tau_B=\exp(-H_B/\kappa_B T)/Z$ is the condensate excitations density matrix assumed to be in equilibrium at temperature T ($\kappa_B$ is the Boltzmann constant and $Z=\textrm{Tr} \exp(-H_B/\kappa_B T)$).

Since we are interested in the two-qubit dynamics only, we trace out the state of the environment and compute the reduced density matrix after time $t$ has passed:
\begin{equation}
\rho(t) = \textrm{Tr}_B\varrho(t).
\end{equation}
The populations of the density matrix do not change with time. This is a consequence of the form of the impurity-condensate density-density interaction which induces pure dephasing in the impurity system.
On the other hand the coherences, represented by the off-diagonal terms, are governed by a non trivial dynamics:
\begin{equation} \label{eq:rhosol}
\rho_{i\neq j}(t) =e^{-\Gamma_{ij}(t) +i \Pi_{ij}(t)} \rho_{i\neq j}(0)
\end{equation}
The quantities $\Gamma_{ij}(t)$ and $\Pi_{ij}(t)$ are real functions of time and describe decay of coherences, that is, proper decoherence, and phase shifts, respectively.
The precise form of these functions was computed in \cite{gabriele} and here we only report the form for the $\Gamma$ factors. Here we focus on a three-dimensional environment.

The off diagonal elements each decay according to one of three decoherence parameters.
The first:
\begin{eqnarray}
	&\Gamma_0(t) = \frac{2g_{AB}^2 n_0}{\pi^2} \int_0^\infty {\text d}k \,k^2 e^{-k^2 \sigma^2/2}  \frac{\sin^2 E_k/2\hbar}{E_k (\epsilon_k + 2 g_B n_0)}  \nonumber \\
&\times \coth \frac{\beta E_k}{2} \left(  1 -\frac{\sin 2 k L}{2 k L} \right), 
\end{eqnarray}
where $\beta=1/(\kappa_B T)$, is responsible for the decoherence of each qubit individually. Therefore it appears in matrix elements such as $\rho_{\{LL\},\{LR\}}$ or $\rho_{\{RL\},\{RR\}}$.

The other two factors entering in matrix elements such as $\rho_{\{LL\},\{RR\}}$ or $\rho_{\{LR\},\{RL\}}$, in which the collective coherence of the two qubits is involved, are
\begin{eqnarray}
	&\Gamma_\pm(t) = \frac{2g_{AB}^2 n_0}{\pi^2} \int_0^\infty {\text d}k\, k^2 e^{-k^2 \sigma^2/2}  \frac{\sin^2 E_k/2\hbar}{E_k (\epsilon_k + 2 g_B n_0)} \coth \frac{\beta E_k}{2} \nonumber \\
&\times \left(  2 -2 \frac{\sin 2 k L}{2 k L} \mp 2 \frac{\sin 2 k D}{2 k D} \pm \frac{\sin 2 k (D+L)}{2 k (D+L)} \pm \frac{\sin 2 k (D-L)}{2 k (D-L)} \right) \nonumber\\
&= 2\Gamma_0(t) \pm \delta(t), \label{eq:Gammapm}
\end{eqnarray}
where $\delta(t)$ represents the cross talk between the qubits via the reservoir. For very short times, due to their spatial separation, each qubit decoheres independently as shown in~\cite{gabriele,us}. For larger times, determined by the ratio of the distance over the typical speed of sound, the three parameters become distinct and the effects of the common reservoir become apparent.  Increasing the distance between the two impurities allows for simulation the dynamics of two qubits in local dephasing environments. Hence, by changing $D$, that is, the wavelength of the superlattice, one can pass from a common environment to an independent environment scenario.  

Starting from Eq.~(\ref{eq:rhosol}) a straightforward calculation allows us to prove that the exact master equation for the system can be written in the following time local form, 
\begin{widetext}
\eqa
\frac{d\rho(t)}{dt}&=&\frac{1}{8}\dot{\Gamma}_+(t)
\left[(\sz^a-\sz^b)\rho(\sz^a-\sz^b)-\frac{1}{2}\{(\sz^a-\sz^b)(\sz^a-\sz^b),\rho\}\right]\nn\\
&+&\frac{1}{8}\dot{\Gamma}_-(t)\left[(\sz^a+\sz^b)\rho(\sz^a+\sz^b)-\frac{1}{2}\{(\sz^a+\sz^b)(\sz^a+\sz^b),\rho\}\right], \label{eq:ME}
\eeqa
\end{widetext}
{where the time-dependent dephasing coefficients are the time derivatives of the decoherence factors given by Eqs.~(\ref{eq:Gammapm}), and where the jump operators are given by $\sz^a \pm \sz^b$. One can verify that the dephasing coefficients may attain negative values, indicating that the dynamics is non-Markovian or, more precisely, that the dynamical map is non-divisible. It is worth stressing that the general conditions under which a non-Markovian master equation can be written in time-local form are not yet known. To the best of our knowledge this is the first example of a two-qubit master equation derived from a microscopic model which is proven to be in time local form.

Our system is a simple example for which the generators of the dynamics form a commutative algebra. For these systems one can generalise the Lindblad-Gorini-Kossakowski-Sudarshan theorem \cite{Lindblad} proving that the dynamics is completely positive if and only if the decoherence factors of Eqs.~(\ref{eq:Gammapm}) are positive at all times \cite{chkos}. The physical implementation and verification of our model would allow then to verify the generalization of the LGKS theorem for the simplest nontrivial commutative open quantum system. Both the single qubit dephasing case and the dephasing of two qubits in local environments, indeed, have been shown to posses a random unitary representation \cite{strunz}.
}

\section{Correlations for initially entangled states} 
Let us consider the dynamics of entanglement between the impurities, and its dependence on the system and reservoir externally controllable parameters.
Due to their computational simplicity combined with their potential for entanglement and non-classical behaviour, we  consider as initial states Werner states of the form
\eq
\label{eq:wernerstates}
\rho_W^{+}\equiv c\ket{\Phi^+}\bra{\Phi^{+}}+\frac{1-c}{4}\mathbb{I}, \quad c\in[0,1],
\eeq
and
\eq
\label{eq:wernerstates2}
\rho_W^{-}\equiv c\ket{\Psi^{+}}\bra{\Psi^{+}}+\frac{1-c}{4}\mathbb{I}, \quad c\in[0,1],
\eeq
where $\ket{\Phi^+}=(\ket{LL}+\ket{RR})/\sqrt{2}$ and  $\ket{\Psi^+}=(\ket{LR}+\ket{RL})/\sqrt{2}$ are maximally entangled states and $\mathbb{I}$ is the 4-dimensional identity matrix. As an example, in Appendix \ref{app} we discuss how to prepare the two qubits in the state given by Eq.~\eqref{eq:wernerstates}. As for the generation of the second class of initial states here considered, given by Eq.~\eqref{eq:wernerstates2}, we note that  all Bell states are equivalent up to local operations, i.e., if we are able to create one then we can generate all others, therefore the procedure sketched in Appendix \ref{app} can be generalised accordingly. Werner states have applications in quantum information and quantum teleportation~\cite{werner1,werner2}. Additionally, they interpolate between maximally entangled and separable states, that is, they can have any value of concurrence between 0 and 1, and therefore are ideal states for investigating entanglement dynamics in our system.  For $c \leq 1/3$ the state is separable; above that, the state initially has non-zero concurrence which will time evolve as
\eq
C_W^{\pm}(t)=\text{max}\left\{0,ce^{-\Gamma_{\mp}(t)}-\frac{1-c}{2}\right\}. \label{eq:cwp}
\eeq
This formula shows that, if the system is initially prepared in a Werner state, the interaction with the condensates cannot increase the initial entanglement and in general leads to a loss of concurrence. However, as we will see in the following, by manipulating the system-reservoir parameters we can control entanglement dynamics and maximise the amount of stationary entanglement in the system. From  Eq.~(\ref{eq:cwp}) we also see that, for an initial state of the form $\rho_W^{+}$ ($\rho_W^{-}$) the only decay rate which enters the dynamics is the collective decay rate $\Gamma_-(t)$ ($\Gamma_+(t)$).

%%%%%%%%%%%%%%%%%%%%%%%%%%%%%%
\begin{figure}[t]
\includegraphics[width=1\linewidth]{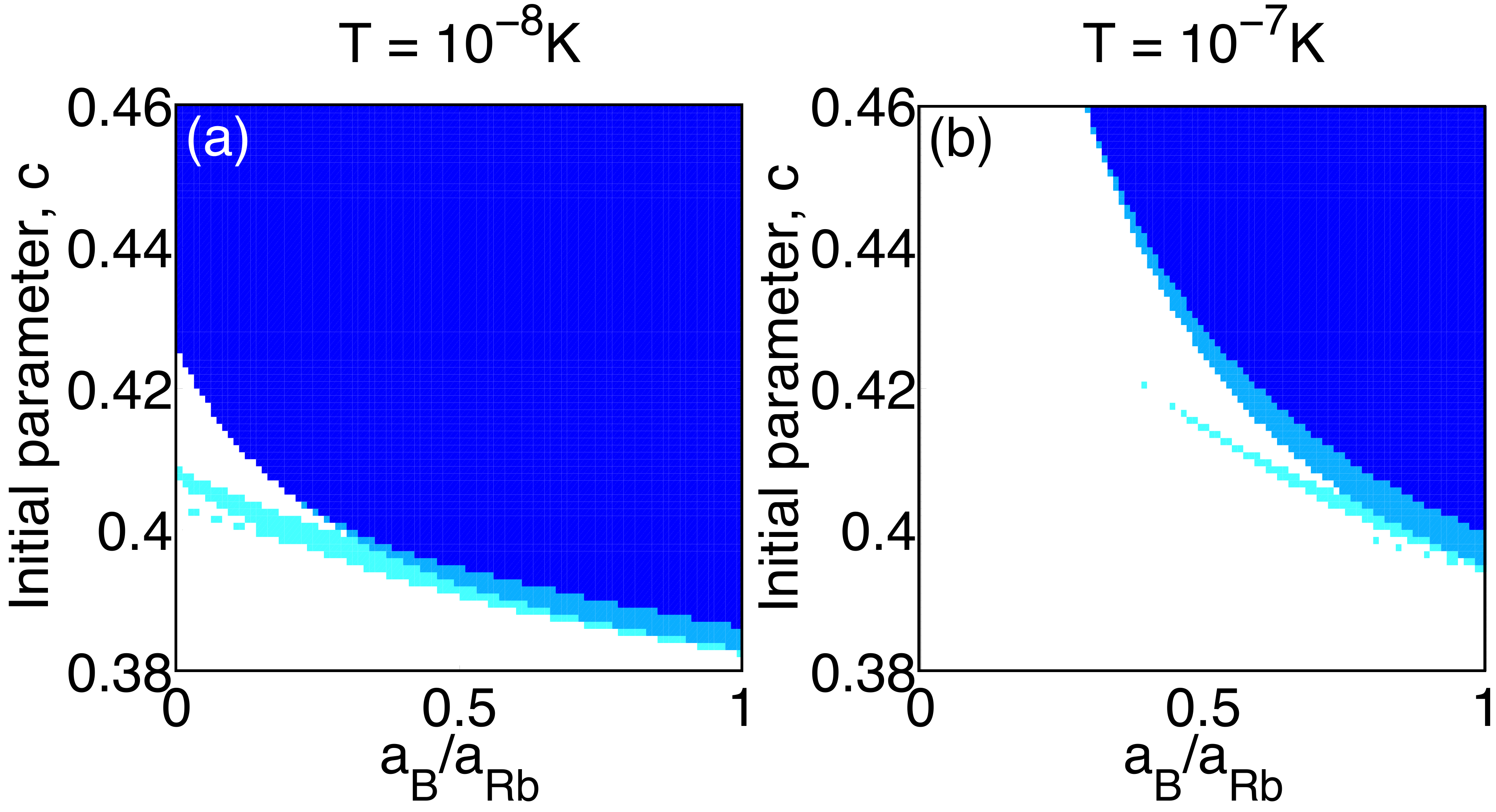}
\caption{(Color online) Entanglement phase diagram for $\rho_W^-$
 as a function of scattering length, $a_B$ in units of $a_{Rb}$, and initial state parameter, $c$, for $D = 5\lambda$ and a) $T = 10^{-8}$K and b) $T = 10^{-7}$K. We distinguish three phases: entanglement trapping (dark blue, top right), entanglement sudden death (white, bottom left) and entanglement revivals (medium blues) (see main text).}
 \label{entphase}
\end{figure}
%%%%%%%%%%%%%%%%%%%%%%%%%%%%%%%

%%%%%%%%%%%%%%%%%%%%%%%%%%%%%%%%%%%%
\begin{figure}[t]
\includegraphics[width=1\linewidth]{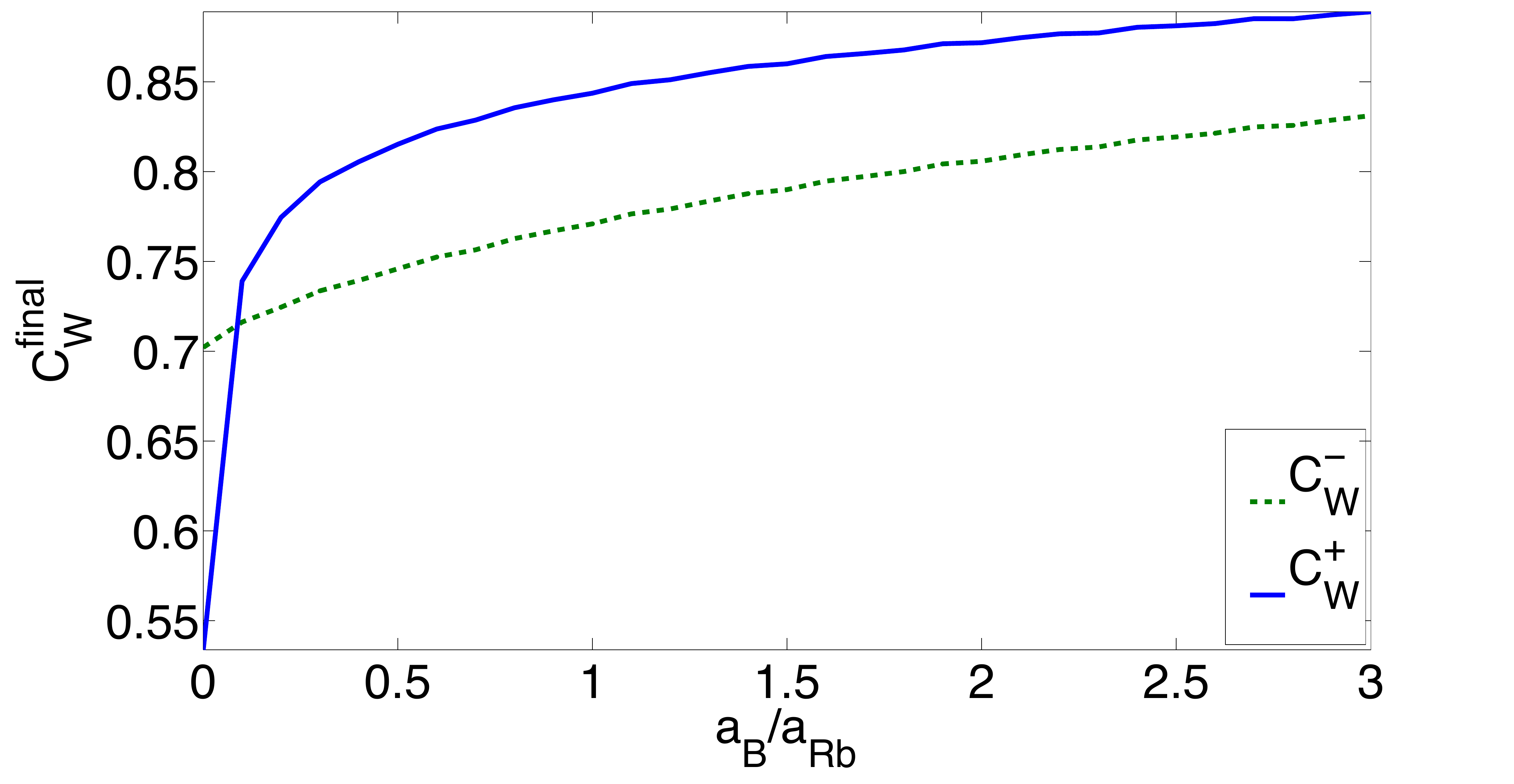}
\caption{(Color online) Stationary concurrence as a function of relative scattering length, $a_B/a_{Rb}$, for  $D = \lambda/2$, $T=10^{-8}$K and for initial states $\rho_W^{-}$ (green dashed line) and $\rho_W^{+}$ (blue solid line).}
\label{statent}
\end{figure}
%%%%%%%%%%%%%%%%%%%%%%%%%%%%%%%%%%%

It is worth noticing that, for the states here considered, $C_W^{\pm}(t)$ does not depend on the phase shifts $\Pi_{ij}(t)$. The behaviour of the concurrence depends primarily on three factors: the initial state parameter, $c$, the scattering length of the condensate, $a_B$, and the distance between the qubits, $D$. By varying these parameters, it is possible to obtain a wide range of entanglement dynamics for long times. Specifically, three different types of entanglement dynamics can be observed: (i) sudden death of entanglement, (ii) sudden death followed by revivals, (iii) entanglement trapping.

Fig.~\ref{entphase}~(a) shows the distribution of the types of entanglement behaviours for two qubits at a distance $D = 5 \lambda$, where $\lambda$ is the wavelength of the optical lattice, for varying initial parameters, $c$, and scattering lengths, $a_B$, measured in units of the natural $^{87}$Rb scattering length $a_{Rb}$, for the initial state of Eq.~(\ref{eq:wernerstates}). The white region corresponds to entanglement sudden death, the medium blue region corresponds to sudden death followed by revivals and the dark blue region corresponds to entanglement trapping. It is worth stressing the extreme sensitivity of the dynamics to the initial state, in the interval $0.38 \lesssim c \lesssim 0.44$. We recall that the state is separable for $c > 1/3 \simeq 0.33$. Entanglement sudden death takes place for small values of initial entanglement, that is, for $0.33 \lesssim c \lesssim 0.38$, while if the initial entanglement is sufficiently high the phenomenon of entanglement trapping, a nonzero value of stationary entanglement, will occur. This is the case for any value of $c \ge 0.425$, independently on the value of the scattering length, hence we can conclude that in this system entanglement trapping is dominant. A similar entanglement phase-diagram exist for initial states of the form of Eq.~(\ref{eq:wernerstates2}).

Entanglement trapping originates from the fact the the time dependent dephasing rates appearing in the master equation~(\ref{eq:ME}) go to zero after a finite time. Hence dephasing stops and so does entanglement loss. This is a strongly non-Markovian phenomenon and never occurs for systems described by master equation with positive constant decay rates.

In between the sudden death and entanglement trapping regions there are two small regions where entanglement exhibits periodic revivals which may or may not result in residual entanglement. It is worth noticing that, even when temperature is increased, there is still entanglement trapping for the majority of cases, as shown in Fig.~\ref{entphase}~(b). The effect of the temperature is to enhance entanglement decay and enlarge the parameter space for which entanglement sudden death occurs. In fact, for typical experimental temperatures of the order of $T =  10^{-8}$K to $T =10^{-7}$K entanglement trapping may only take place for sufficiently strongly interacting gases. In general, an increase in the scattering length of the ultracold reservoir will favor entanglement trapping, allowing for this phenomenon to occur for a wider class  of initial states. Moreover, as we can see from Fig.~\ref{statent}, the stationary entanglement increases for increasing values of $a_B$, for both classes of initial states. This conclusion holds for any value of the distance $D$ between the impurities. Hence we conclude that more strongly interacting gases are optimal for reaching higher values of entanglement trapping.

%%%%%%%%%%%%%%%%%%%%%%%%%%%%%%%
\begin{figure}[t]
\includegraphics[width=1\linewidth]{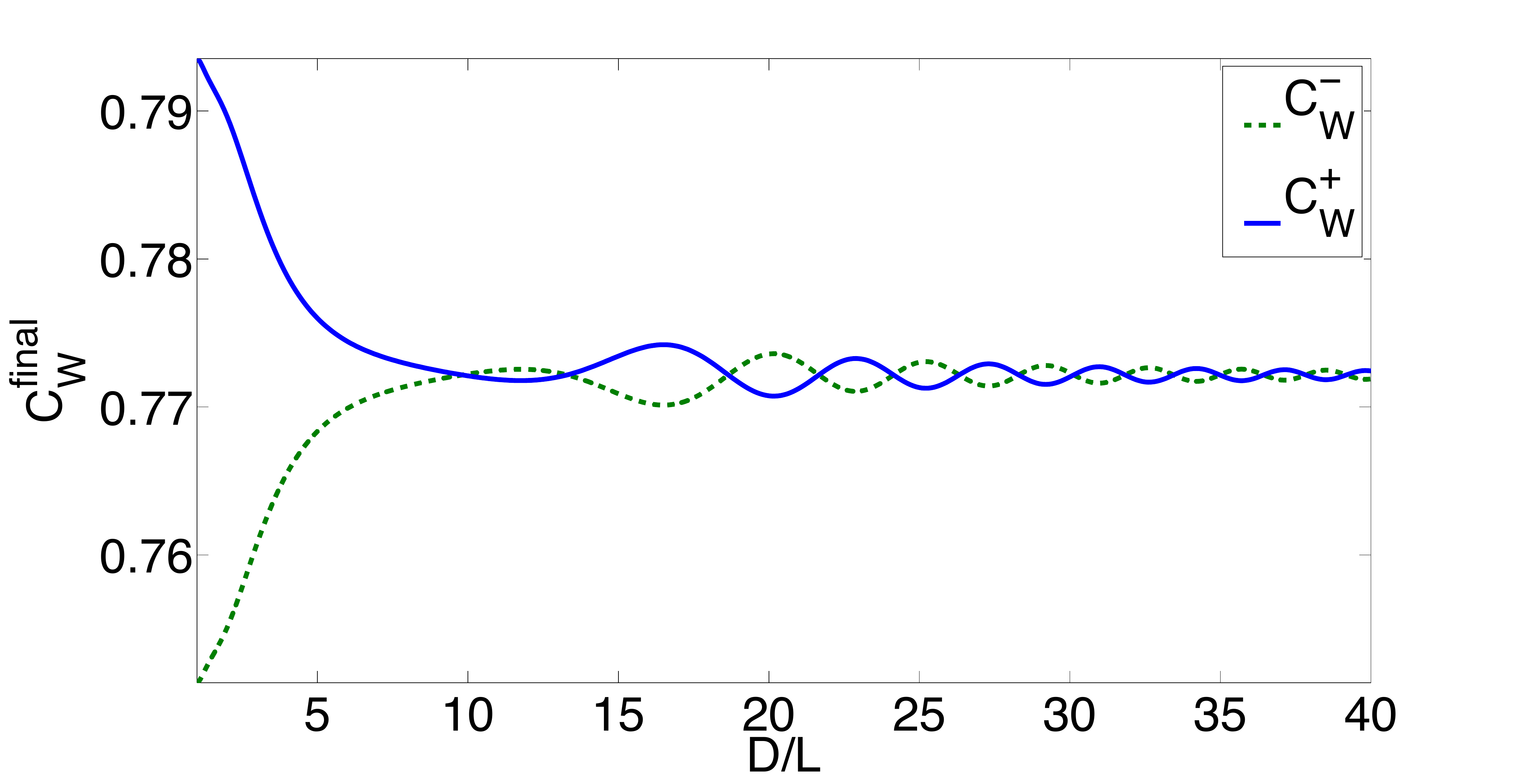}
\caption{(Color online) Steady state value of concurrence, $C_W^{final}$, as a function of distance between qubits for initial states $\rho_W^+$ (blue solid) and $\rho_W^-$ (green dashed) for $a_B = a_{Rb}$.}
\label{final_vs_D}
\end{figure}
%%%%%%%%%%%%%%%%%%%%%%%%%%%%%%%

We now investigate the change in stationary entanglement when increasing the qubits separation, for a fixed value of the scattering length. Figure~\ref{final_vs_D} shows $C_W^-$ and $C_W^+$ as a function of $D/L$. Initially, there is an increase or decrease, respectively, of the entanglement, settling towards a steady value as the qubits get further apart. The different initial behaviour of stationary entanglement for the two classes of initial states can be explained as follows. When immersed in a common environment the initial states $\rho_W^{+}$, which are mixtures of the subradiant state $\ket{\Phi^{+}}$, are sub-decoherent \cite{gabriele}. In our model an increase in the distances corresponds to a vanishing effect of the cross talk term, $\delta$, and hence to a transition to a model of local dephasing for the impurities. For locally dephasing impurities sub-decoherence does not occur. This explains why an increase in the distance causes a decrease in the trapped entanglement for initially sub-decoherent states. The opposite holds for the other class of initial states $\rho_W^{-}$, which are super-decoherent in a common environment. In general, for large $D$, the cross talk term, $\delta$, vanishes and the qubits behave as single qubits in independent reservoirs, eliminating the difference caused by their initial states. 
In terms of reservoir engineering, states $\rho_W^{+}$ are best protected from decoherence when immersed in a common environment, that is, for shorter values of the impurity separation due to smaller superlattice wavelength, while states $\rho_W^{-}$ maximally retain the initial entanglement for longer impurity separations, that is, larger superlattice wavelengths.

\begin{figure}[t]
\includegraphics[width=1\linewidth]{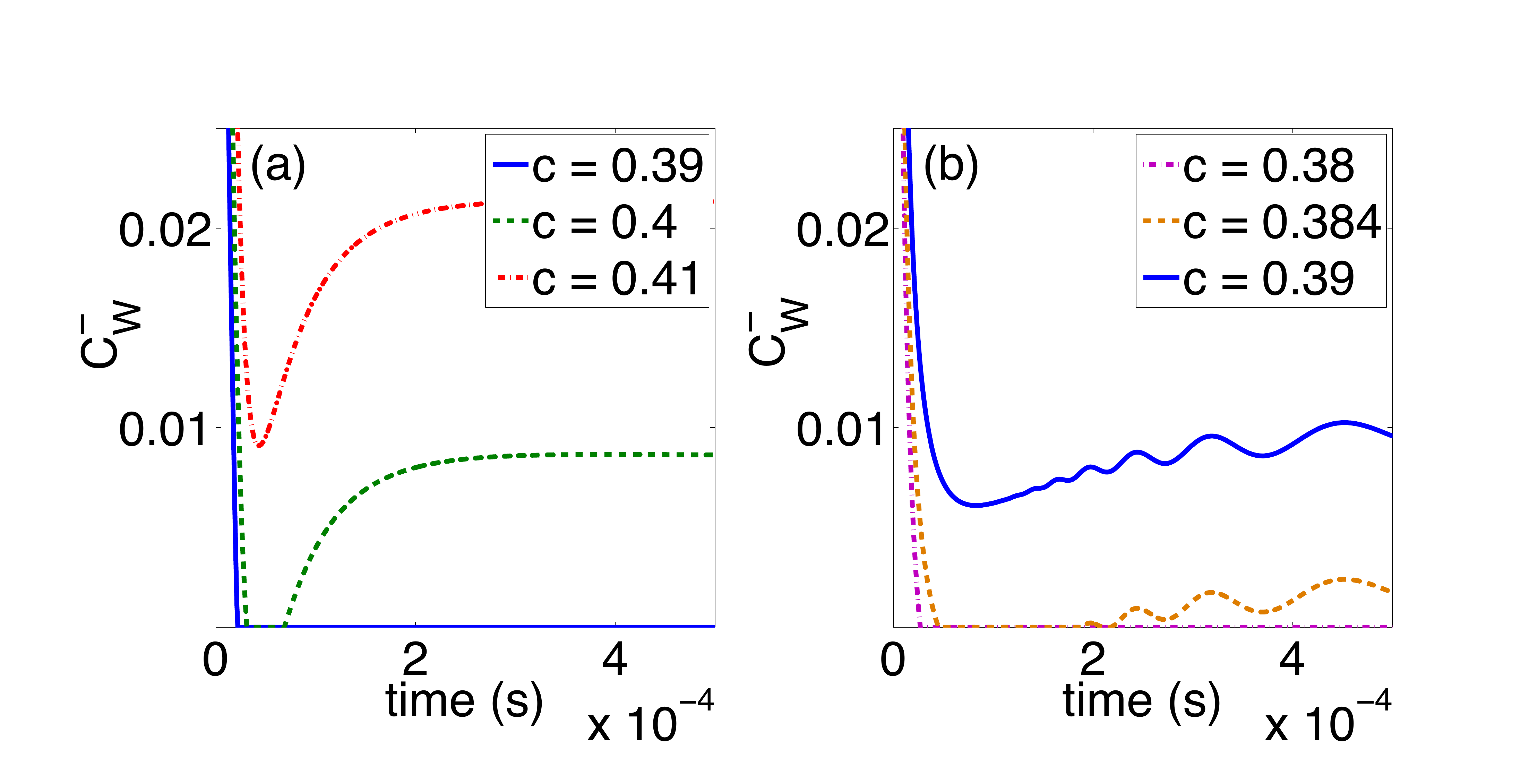}
\caption{(Color online) Concurrence, $C_W^+$, for  a) $D = \lambda/2$ for initial state parameters $c = 0.34$ (blue solid), $0.341$ (green dashed) and $0.342$ (red dotted), and  b) $D = 5\lambda$ for initial state parameters $c = 0.339$ (purple dotted), $0.3395$ (orange dashed) and $0.34$ (blue solid).
}
\label{concurrences}
\end{figure}

Until now, we have analysed the reservoir parameters maximising the value of residual entanglement. We now look at the changes in the entanglement dynamics for different initial states and impurity distances. We focus our attention on the most interesting region of system-reservoir parameter space where the entanglement is most sensitive to the initial conditions.  Fig.~\ref{concurrences}~(a) shows the concurrence for two qubits at $D = 2L = \lambda/2$, that is, in adjacent pairs of sites of an optical lattice, for three values of the initial state parameter, $c$. As $c$ increases, we can see the concurrence move from zero value to a finite value in the steady state. Fig.~\ref{concurrences}~(b) shows the same for qubits located at $D = 5\lambda$. At these distances the cross talk term, $\delta$, picks up oscillations and the resulting concurrence shows more varied behaviour. In addition, the same initially entangled state will result in different concurrences for different values of $D$. In particular, for the same initial state, $c = 0.34$, the steady state value of the concurrence changes from zero in Fig.~\ref{concurrences}~(a) to a small but non-zero value for Fig.~\ref{concurrences}~(b) due to the increased distance between qubits.

%%%%%%%%%%%%%%%%%%%%%%%%%%%%%%%%%%%%
%%%%%%%%%%%%%%%%%%%%%%%%%%%%%%%%%%%

\begin{figure}[h]
\includegraphics[width=1\linewidth]{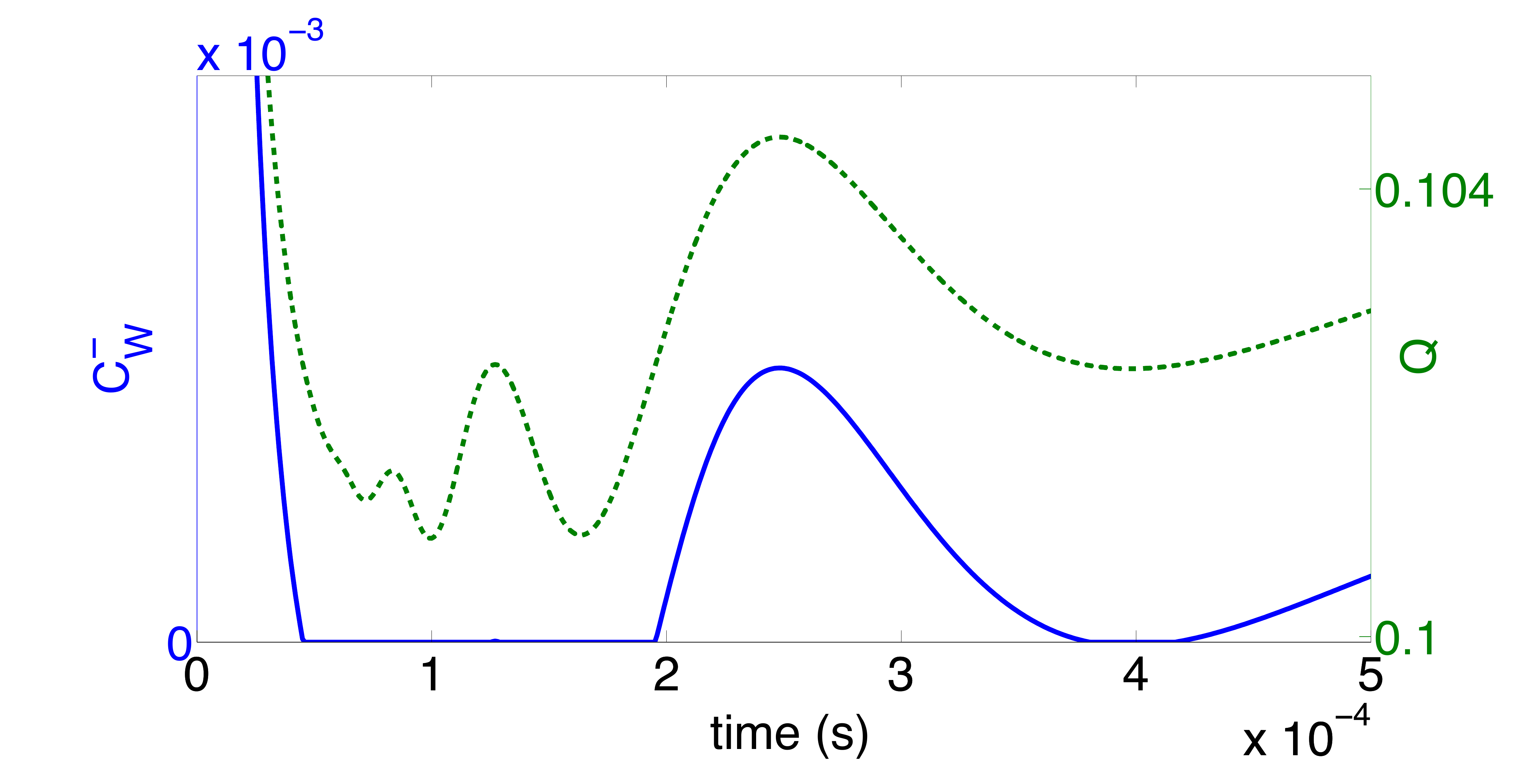}
\caption{(Color online) Concurrence for the initial Werner state (blue solid) and discord (green dashed) as a function of time for $a_B = a_{Rb}$ and $D = 10L$ for initial parameter $c = 0.3395$.}
\label{discord}
\end{figure}

For the sake of completeness, we conclude our analysis of correlations dynamics by looking at the quantum discord of the impurities, as an example of a quantum correlation that exceed those captured by entanglement \cite{discord}. We consider the original definition of discord:  
\eq
	\cal{Q}(\rho) = \cal I (\rho) - \cal C (\rho),
\eeq
where $\cal I(\rho)$ is the quantum mutual information and $\cal C(\rho)$ quantifies the classical correlations. As such, this quantity describes correlations beyond those ascribable to classical physics and highlights the presence of quantum effects even in cases where there is no entanglement. For Werner states of the form outlined above, the discord can be calculated using a simple algorithm~\cite{discordcalc}. In Fig.~\ref{discord} we compare the discord and concurrence as a function of time. Here the concurrence shows an initial loss of entanglement, followed by periods of revival. The behaviour of the discord follows the same general trends as the concurrence, having peaks for the same instants of time. For this class of states, this is probably due to the simple dependance of both quantities on qubits correlations $\langle \sigma_z \sigma_z \rangle$ and $\langle \sigma_x \sigma_x \rangle$.
An important difference, however, is that in the time intervals in which entanglement temporarily disappears, there are still oscillations present in the discord showing that the state still displays non-classical behaviours. This is not surprising as, in general, it is known that quantum discord is more robust against decoherence than entanglement and, for example, does not exhibit the phenomenon of entanglement sudden death.

%%%%%%%%%%%%%%%%%%%%%%%%%%%%%%%
%%%%%%%%%%%%%%%%%%%%%%%%%%%%%%%

\section{Initial product states and entanglement generation} 
\begin{figure}[h]
\includegraphics[width=1\linewidth]{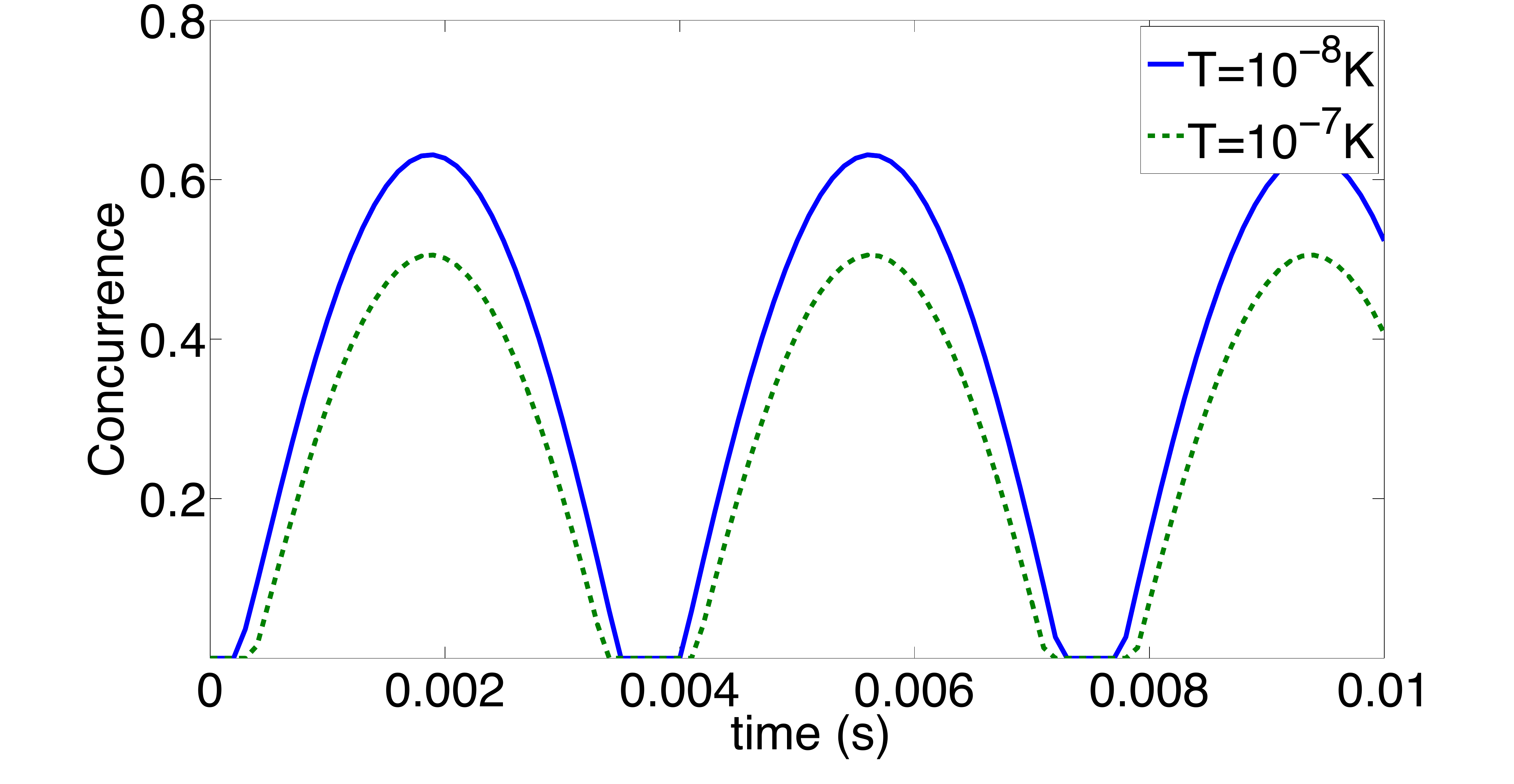}
\caption{(Color online) Concurrence for the initial state of Eq.~(\ref{eq:initialstate}) as a function of time for $T = 10^{-8}$K (solid blue) and $T = 10^{-6}$K (green dashed).}
\label{ent_gen}
\end{figure}

In the previous section, we have considered the effect of the environment on states that are initially entangled and have shown that the presence of a carefully chosen shared environment protects the system from loss of entanglement. However, one key advantage of a shared environment is its ability to generate correlations between initially separable states. In analogy to \cite{braun}, here we consider an initial product state,
\eq
\label{eq:initialstate}
	\ket{\psi(0)} = \frac 12 \left(\ket{LL}+\ket{LR}+\ket{RL}+\ket{RR} \right),
\eeq
and evolve it in time. In contrast to Werner states in which only one of the decay rates, $\Gamma_\pm(t)$, would appear, for this initial state the dynamics is dictated by all three decay rates, including $\Gamma_0$. Moreover in this case the entanglement crucially depends on the phase shifts $\Pi_{ij}(t)$. In order to measure any generated non-classical correlations developed in the system, we calculate the concurrence of the resulting state. Fig.~\ref{ent_gen} shows the concurrence as a function of time for this initial state. The concurrence shows a periodic oscillation between its zero value and a value close to its maximum $1$, for temperatures of $T = 10^{-8}$K. If the temperature is increased to $T = 10^{-6}$K, the periodic generation of entanglement remains, but the maximum attainable concurrence is reduced.

The origin of this entanglement and the specific choice of the particular initial state \eqref{eq:initialstate} can be explained by the fact that the BEC induces an effective interaction Hamiltonian proportional to $\sigma_z\sigma_z$. This interaction creates conditional shifts between the two qubit impurities which, for a state of the form \eqref{eq:initialstate}, correspond to an entangling operation. Therefore for other product states such as $\ket{LL}$ or $\ket{LR}$ we do not observe any entanglement generation at all.

\begin{figure}[h]
\includegraphics[width=1\linewidth]{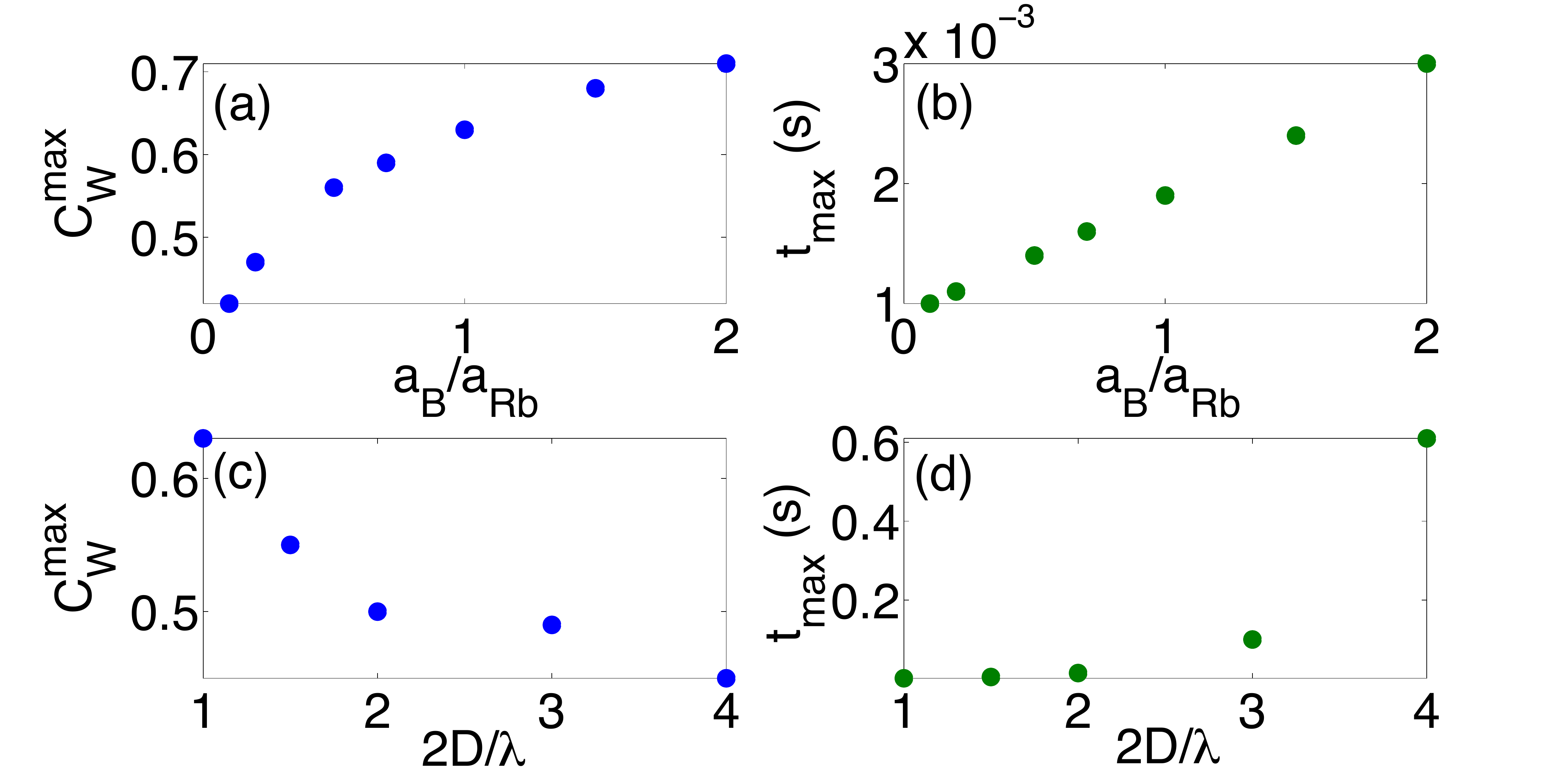}
\caption{(Color online) a) $C_W^{max}$, the maximum concurrence generated for the initial state of Eq.~(\ref{eq:initialstate}), as a function of scattering length $a_B/a_{Rb}$ for $D = \lambda/2$ and $T = 10^{-8}$K. b) $t_{max}$, the time at which the maximum concurrence of (a) occurs.\\
c) $C_W^{max}$, the maximum concurrence generated for the initial state of Eq.~(\ref{eq:initialstate}), as a function of distance between qubits, $D$ for $a_B = a_{Rb}$ and $T = 10^{-8}$K. d) $t_{max}$, the time at which the maximum concurrence of (c) occurs. }
\label{ent_gen2}
\end{figure}

We now explore how the generated entanglement can be maximised by varying reservoir parameters. Fig.~\ref{ent_gen2} shows the dependence of both the maximal entanglement  and the generation time when one changes either the scattering length $a_B$ or the distance $D$ between the impurities. As the entanglement generation crucially relies on the phase factors, and as these quantities have negligible effect for increasing distances $D$, one sees that the maximum concurrence decreases with increasing distance while the generation time increases. On the other hand, the plots show that a more strongly interacting BEC leads to higher values of entanglement, that is, once more the optimal reservoir configuration is for higher values of $a_B$. The price to pay is a greater value of the generation time, indicating that one has to wait longer for the system to attain the highly entangled state.

Apart from fundamental implications for quantum reservoir engineering, our setup may have applications in optical lattices for the creation of long distance entanglement between impurities trapped in a superlattice. One does not need to manipulate the form of the impurities' potential by, for example, lowering and raising the barrier in a double well. One needs to control the time interval that impurities and BEC interact by moving the impurities away from the BEC or by lowering to zero the interaction strength $g_{AB}$ by means of Feshbach resonances.

%%%%%%%%%%%%%%%%%%%%%%%%%%%%%%%

%%%%%%%%%%%%%%%%%%%%%%%%%%%%%%%
%%%%%%%%%%%%%%%%%%%%%%%%%%%%%%%

\section{Conclusions}
 Atoms immersed in Bose-Einstein condensates provide an ideal system for investigating and probing many-body dynamics thanks to their interaction with the quantum excitations of the condensate. We have shown that the combination of the double well qubits and a Bose-Einstein condensate reservoir offers several advantages for quantum information processing and storage. For initially entangled Werner states, there is an initial loss of entanglement to the environment, but the presence of the common environment reduces, and in some cases partly reverses, the loss of correlations from the system to the environment. Moreover, for certain initially separable states, the BEC-induced effective interaction can generate entanglement between the two distant qubits.  In the spirit of reservoir engineering, we have demonstrated how we can manipulate both the entanglement dynamics and the residual entanglement by modifying system-reservoir parameters and we have shown that more strongly interacting BECs are optimal for both entanglement generation and entanglement trapping. Moreover, we have shown that the dependence on the distance between impurities is non trivial. If one aims at prevention of entanglement, then different classes of initial Werner states have opposite requirements ($D/L \simeq 1$ for $\rho_W^+$ and $D/L \gg 1$ for $\rho_W^-$), while entanglement generation is always optimal for small distances.
 
Finally, we have shown that our results are robust to the effects of experimentally realistic temperatures, meaning that our proposal may be tested in present-day experimental setups. Our results pave the way to  implementations of quantum communication protocols in arrays of quantum impurities and underline at the same time the potential of these systems as testbeds for fundamental studies on open quantum systems in the non-Markovian regime.

\begin{acknowledgments}
This work was supported by EPSRC (EP/J016349/1), the Emil Aaltonen foundation, and the Finnish Cultural foundation. We acknowledge Markus Cirone for useful discussions.
\end{acknowledgments}

\appendix
\section{Experimental realization}
\label{app}
The ideas and schemes presented in this work can be implemented using techniques realised in recent experiments of impurity-BEC dynamics. We consider an implementation with ${}^{133}$Cs impurity atoms immersed in a ${}^{87}$Rb condensate, but our results can be extended to other species or a single specie BEC but with two different internal states. The impurities can be trapped by an optical superlattice with a double well elementary cell. In the simplest case $D=2L=\lambda/2$. For the case described in Fig.~\ref{final_vs_D} in which $L$ and $D$ are not in a simple ratio, one needs to use a different trapping mechanism for the impurities, for example arrays of microtraps \cite{birkl} or of optical tweezers \cite{kuhn}. 

The initial Werner states considered in Eq.~\eqref{eq:wernerstates} can be created using an entangling operation as the ``square root of swap"  operation realized in \cite{anderlini} for two internal states $\ket{0}$ and $\ket 1$ of the two atoms. 
The square root of swap gate is defined as:
\begin{equation}
\sqrt{\textrm{SWAP}} = \left(\begin{array}{cccc}
1 & 0 & 0 & 0 \\
0 & 1/\sqrt 2 & 1/\sqrt 2 & 0 \\
0 & -1/\sqrt 2 & 1/\sqrt 2 & 0 \\
0 & 0 & 0 & 1
\end{array}\right)
\end{equation}
The protocol to create the Werner state $\rho_W^+$ is as follows: i) prepare the two atoms in two distinct single wells; with probability $c$ create the state $\ket{10}$ and with probability $1-c$ create the maximally mixed state of the two qubits $\mathbb{I}/4$ by applying a fast random phase to an equal superposition of the two states $\ket 0$ and $\ket 1$ for the two atoms; ii) apply the $\sqrt{\textrm{SWAP}}$ gate by bringing the two atoms together, let them interact and separate them again; iii) finally transfer the internal state of each atom into the left and right states of the double well, that is, $\ket 0\to\ket L$ and $\ket 1\to\ket R$ using a spin dependent potential and at the same time transforming each single well into a double well by raising the central barrier. To create the state $\rho_W^-$ use $\ket{01}$ in stage i).
A similar procedure, not involving the $\sqrt{\textrm{SWAP}}$ can be used to create state Eq.~\eqref{eq:initialstate}.

Finally the readout can be achieved by transferring back the positional states $\ket L$ and $\ket R$ into the internal states $\ket 0$ and $\ket 1$ and then doing a full tomography of the two-qubit state.


\begin{thebibliography}{30}
\bibitem{note} The bosonic or fermionic nature of the impurity and thus the commutation relations fulfilled by $\hat\Psi$ are irrelevant for our model as there is only one impurity per double-well.
\end{thebibliography}
\end{document}